\documentstyle[aps,preprint]{revtex}
\begin{document}
\draft
\preprint{ISSP \today}
\title{Exact Results on Superconductivity
due to Interband Coupling}
\author{Y. Morita$^*$$^{1}$, Y. Hatsugai$^{**}$$^{2}$ and M. Kohmoto$^{1}$ }
\date {\today}
\address{
 $^{1}$Institute for Solid State Physics,
 University of Tokyo,
 7-22-1 Roppongi Minato-ku, Tokyo 106, Japan
 }
\address{
 $^{2}$Department of Applied Physics,
University of Tokyo, 7-3-1 Hongo, Bunkyo-ku, Tokyo 113, Japan}
\maketitle
\begin{abstract}
We present a family of exactly solvable models
at arbitrary filling in any dimensions which exhibit
novel superconductivity with interband pairing.
By the use of the hidden $SU(2)$ algebra
the Hamiltonians were diagonalized explicitly.
The zero-temperature phase diagrams
and the thermodynamic properties are discussed.
Several new properties are revealed
which are different from those of the BCS-type superconductor.
\end{abstract}
\narrowtext

Superconductivity is one of the most remarkable phenomena
in condensed matter physics.
Recently possibilities of a novel superconductor
are proposed by Kohmoto and Takada $\cite{KT1}$.
They investigated the superconducting instability of insulators
by the mean-field treatment.
A two-band system which is insulating
without interactions becomes superconducting
by a sufficiently large interband attraction.
It has many properties
which are different
from those of the BCS-type superconductors $\cite{BCS}$.
Note that the Cooper instability is irrelevant here,
since there is no Fermi surface.
In Ref.$\cite{KT2}$, possible realization in organic materials
is discussed,
which is an extension of the Little's idea
for the room-temperature superconductor $\cite{Little}$.

We have constructed a family of exactly solvable models
at arbitrary filling in any dimensions
which includes the models proposed in Ref.$\cite{KT1}$ and Ref.$\cite{KT2}$.
We have obtained the ground state
and the thermodynamic quantities explicitly.
Several new properties have been revealed.
An instability without a Fermi surface,
which was proposed by Kohmoto and Takada,
is realized in the models.
This instability is quite different from the Cooper instability.
A finite strength of attraction
is needed to produce the superconductivity
in contrast to the BCS-type superconductivity.

Let us consider a two-band model described by the Hamiltonian
\begin{eqnarray}
{\cal H}&=& {\cal H}_{\rm kin}+{\cal H}_{\rm int},
\label {hamiltonian1}
                       \\
{\cal H}_{\rm kin} &=& \sum_{k}
          \epsilon ^{(v)} (k) c_{k}^{(v)^{\dagger}} c_{k}^{(v)}
         +\sum_{k} \epsilon ^{(c)} (k) c_{k}^{(c)^{\dagger}}
        c_{k}^{(c)},
\label {hamiltonian11}
                        \\
{\cal H}_{\rm int} &=&  -\frac {U}{N}
             (\sum _{k} c_{k}^{(c)^{\dagger}}
             c_{-k}^{(v)^{\dagger}})
             (\sum _{k} c_{-k}^{(v)}
              c_{k}^{(c)}),
\label {hamiltonian12}
\end{eqnarray}
where $c_{k}^{(v)}$ and $c_{k}^{(c)}$
are
the fermion annihilation operators
for the valence band and the conduction band
and
$\epsilon ^{(v)}(k)$ and $\epsilon ^{(c)} (k)$
are
the energy dispersions of the valence band and the conduction band,
respectively.
The momentum vector $k$ takes values
in the $d$-dimensional Brillouin zone.
We impose a constraint ``symmetric condition'' on the band structure
\begin{equation}
\epsilon ^{(v)}(k)+\epsilon ^{(c)}(-k)=C,
\label {constraint}
\end{equation}
where $C$ is independent of $k$.
Without loss of generality we set $C=0$.
We set $U$ positive and $\sim O(N^0)$,
where $N$ is the number of the momentum points
in the Brillouin zone.
The interaction
is an interband attraction.
The spin degrees of freedom are neglected for simplicity,
since we do not consider the spin-related quantities here.

Let us sketch the process of the diagonalization.
The diagonalization consists of two steps.
At first we show the ``decoupling property'' of the Hamiltonian.
Next we map the system
to an exactly solvable quantum spin system.
Then we can construct all the eigenvalues and the eigenvectors.

Represent
the states in the Hilbert space diagrammatically ( see Fig. 1 ).
Let us span the Hilbert space
by the base vectors
\begin{eqnarray}
&&\ c_{-{p}_{1}}^{(v)^{\dagger }}
\cdots
c_{-{p}_{N_{e}^{v}}}^{(v)^{\dagger }}
c_{{q}_{1}}^{(c)^{\dagger }}
\cdots
c_{{q}_{N_{e}^{c}}}^{(c)^{\dagger }}
\ |0 \rangle _{{\cal S}}
\otimes
(c_{-k_{1}}^{(v)^{\dagger }}c_{k_{1}}^{(c)^{\dagger }})
(c_{-k_{2}}^{(v)^{\dagger }}c_{k_{2}}^{(c)^{\dagger }})
(c_{-k_{3}}^{(v)^{\dagger }}c_{k_{3}}^{(c)^{\dagger }})
\cdots
(c_{-k_{M}}^{(v)^{\dagger }}c_{k_{M}}^{(c)^{\dagger }})
\ |0 \rangle _{{\cal D}},
\label {basevectors}
\end{eqnarray}
where
$\{{p}_{1},\cdots ,{p}_{N_{e}^{v}},
{q}_{1},\cdots ,{q}_{N_{e}^{c}} \}
={\cal S}$
and
$k_{1}, k_{2}, k_{3}, \cdots , k_{M}\in {\cal D}$
($\ $The sets ${\cal S}$ and ${\cal D}$ will be defined below$\ $).
Here $|0 \rangle _{{\cal S}}$ is defined by
$c_{k}^{(c)}|0 \rangle _{{\cal S}} =0\ (k\in {{\cal S}})$
and
$c_{k}^{(v)}|0 \rangle _{{\cal S}} =0\ (-k\in {{\cal S}})$.
$|0 \rangle _{{\cal D}}$ is defined by
$c_{k}^{(c)}|0 \rangle  _{{\cal D}}=0\ (k\in {{\cal D}})$
and
$c_{k}^{(v)}|0 \rangle _{{\cal D}}=0\ (-k\in {{\cal D}})$.
Consider a pair which consists of
the momentum point $-k$ in the valence band
and
the momentum point $k$ in the conduction band.
We denote the pair by $k$,
where $k$ takes values in the Brillouin zone.
Define the sets ${\cal S}$ and ${\cal D}$ as follows.
If $k$ is single-occupied,
$k$ belongs to ${\cal S}$.
And, if $k$ is empty or double-occupied,
$k$ belongs to ${\cal D}$.
Note
${\cal S}\cap {\cal D}=\phi $
and
${\cal S}\cup {\cal D}=$the Brillouin zone.

Let us introduce an operator ${\cal P}^{j}$
which is a projection operator to the Hilbert space
where  ${\cal S}$ and ${\cal D}$ are fixed
to be ${\cal S}_{j}$ and ${\cal D}_{j}$.
The index $j$ denotes how ${\cal S}$ and ${\cal D}$
are fixed.
Using the properties of ${\cal P}^{j}$,
rewrite the Hamiltonian as
\begin{eqnarray}
{\cal H} &=& ({\sum }_{j}{\cal P}^{j})\ {\cal H}\ ({\sum }_{j}
              {\cal P}^{j})
         \nonumber
         \\
         &=& {\sum }_{j}{\cal P}^{j}{\cal H}{\cal P}^{j}.
\label {projection}
\end{eqnarray}
Using the relation (\ref {constraint}), we have
\begin{eqnarray}
{\cal P}^{j}{\cal H}{\cal P}^{j}
&=&
{\cal P}^{j}{\cal H_{\rm kin}}{\cal P}^{j}
+
{\cal P}^{j}{\cal H_{\rm int}}{\cal P}^{j}
\nonumber
                        \\
&=&
{\cal P}^{j}({\cal H_{\rm I}}\otimes {\bf 1}){\cal P}^{j}
+
{\cal P}^{j}({\bf 1}\otimes {\cal H_{\rm II}}){\cal P}^{j},
\label {mapping}
\end{eqnarray}
where {\bf 1} is an identity operator
and ${\cal H_{\rm I}}$ and  ${\cal H_{\rm II}}$
are
\begin{eqnarray}
{\cal H}_{\rm I}
&=&
\sum_{k \in {\cal S}_{j}}
\epsilon ^{(v)} (-k) c_{-k}^{(v)^{\dagger}} c_{-k}^{(v)}
+\sum_{k \in {\cal S}_{j}} \epsilon ^{(c)} (k)
c_{k}^{(c)^{\dagger}} c_{k}^{(c)},
\nonumber
                        \\
{\cal H}_{\rm II}
&=&  -\frac {U}{N}
(\sum _{k\in {\cal D}_{j}} c_{k}^{(c)^{\dagger}} c_{-k}^{(v)^{\dagger}})
(\sum _{k\in {\cal D}_{j}} c_{-{k}}^{(v)} c_{{k}}^{(c)}).
\label {hamiltonian21}
\end{eqnarray}
Here,
the kinetic term and the interaction term
decouple,
the ``decoupling property'' of the Hamiltonian.

Now we  map the system to an exactly solvable quantum spin system
( see Fig. 1 ).
Here the $SU(2)$ algebra
hidden in spinless fermions in a two-band system
plays a crucial role $\cite {neta1}$ $\cite {neta2}$ $\cite {neta3}$.
Let us define the ``spin'' operators
${\hat S}_{k}^{+}={\hat S}_{k}^{x}+i{\hat S}_{k}^{y}$,
${\hat S}_{k}^{-}={\hat S}_{k}^{x}-i{\hat S}_{k}^{y}$
and
${\hat S}_{k}^{z}$
by
${\cal P}^{j}c_{-k}^{(v)}c_{k}^{(c)}{\cal P}^{j}$,
${\cal P}^{j}c_{k}^{(c)^{\dagger}} c_{-k}^{(v)^{\dagger}}{\cal P}^{j}$
and
${\cal P}^{j}(\frac {1}{2}-c_{k}^{(c)^{\dagger}} c_{-k}^{(v)^{\dagger}}
c_{-k}^{(v)}c_{k}^{(c)}){\cal P}^{j}$ respectively,
the ``total spin'' operators
${\hat S}^{\alpha }$ by
$\sum _{k\in {\cal D}_{j}}{\hat S}_{k}^{\alpha }
\ (\alpha =x, y, z)$
and
$({\bf {\hat {S}}})^{2}$
by
$({\hat S}^{x})^{2}+({\hat S}^{y})^{2}+({\hat S}^{z})^{2}$.
Then we have
\begin{equation}
{\cal P}^{j}({\bf 1}\otimes {\cal H_{\rm II}}){\cal P}^{j}
 =
{\cal P}^{j}({\bf 1}\otimes {\cal H_{\rm spin}}){\cal P}^{j},
\label{spinmapping}
\end{equation}
where $\cal H_{\rm spin}$ is defined by
\begin{eqnarray}
{\cal H}_{\rm spin}
&=&
-\frac {U}{N}
(\sum_{k\in {\cal D}_{j}}
{\hat S}_{k}^{x}-i{\hat S}_{k}^{y})
(\sum_{k\in {\cal D}_{j}}
{\hat S}_{k}^{x}+i{\hat S}_{k}^{y})
\nonumber
                        \\
&=&
-\frac {U}{N}
({\hat S}^{x}-i{\hat S}^{y})
({\hat S}^{x}+i{\hat S}^{y})
\nonumber
                        \\
&=&
-\frac {U}{N}
\{ ({\bf {\hat {S}}})^{2}
-({\hat S}^{z})^{2}-({\hat S}^{z})\}.
\label {hamiltonian22}
\end{eqnarray}
The operators defined above
satisfy the relations
\begin{equation}
[\ S_{k}^{l},S_{\tilde k}^{m}\ ]
=i\epsilon _{lmn}S_{k}^{n}\delta _{k{\tilde k}},
\label {su21}
\end{equation}
\begin{equation}
({\hat S}^{x}_{k})^{2}+({\hat S}^{y}_{k})^{2}+({\hat S}^{z}_{k})^{2}
=\frac {1}{2}\ (\ \frac {1}{2}+1\ ),
\label {su22}
\end{equation}
where $k$ and $\tilde k$ take values in ${\cal D}_{j}$.
Thus ${\hat S}_{k}^{x}, {\hat S}_{k}^{y}\  {\rm and}\  {\hat S}_{k}^{z}$
($k\in {\cal D}_{j}$) are the components of a $s=\frac {1}{2}$ quantum spin.
Now we can identify $k$ with a ``site''
on which a $s=\frac {1}{2}$ quantum spin is defined.
In the language of spin,
if the pair $k$ is empty,
the spin on the site $k$ is ``up''
and
if the pair $k$ is double-occupied,
the spin on the site $k$ is ``down''.
Note that, since $k$ takes values in ${\cal D}_{j}$,
all the pairs we now consider are either empty or double-occupied.
Now diagonalize
${\cal H}_{\rm spin}$
which can be identified
with the Hamiltonian of the quantum spin system ($\ s=\frac {1}{2}\ $).
Define $\ |\ S, \ S^{z}\ \rangle$
by an eigenstate of $({\bf {\hat {S}}})^{2}$ and ${\hat S}^{z}$
which satisfies
$\ ({\bf {\hat {S}}})^{2}\ |\ S, \ S^{z}\ \rangle
   =S\ (\ S+1\ ) \ |\ S, \ S^{z}\ \rangle\ $
and
$\ {\hat S}^{z}\ |\ S, \ S^{z}\ \rangle
   =S^{z}\ |\ S, \ S^{z}\ \rangle\ $.
The energy is specified by $S$ and $S^{z}$ ( see (${\ref {hamiltonian22}}$) ).
There is, however, non-trivial degeneracy which is given by
$\frac {{(2S_{\rm max})}\ !\ (2S+1)}
 {(S_{\rm max}-S) !\ (S_{\rm max}+S+1)\ !}$
where $N_{{\cal D}_{j}}$ is the number of elements in ${\cal D}_{j}$
and $S_{\rm max}$ is $\frac {N_{{\cal D}_{j}}}{2}$.
This degeneracy is crucial for the thermodynamic properties.
Let us consider the state
\begin{equation}
c_{-{p}_{1}}^{(v)^{\dagger }}
\cdots
c_{-{p}_{N_{e}^{v}}}^{(v)^{\dagger }}
c_{{q}_{1}}^{(c)^{\dagger }}
\cdots
c_{{q}_{N_{e}^{c}}}^{(c)^{\dagger }}
\ |\ 0\ \rangle
\otimes
\ |\ S, \ S^{z}\ \rangle ,
\label {eigenvector}
\end{equation}
where $\{{p}_{1},\cdots ,{p}_{N_{e}^{v}},
{q}_{1},\cdots ,{q}_{N_{e}^{c}} \}
={{\cal S}_{j}}$.
{}From the ``decoupling property'' ($\ref {mapping}$)
and the mapping to the quantum spin system ($\ref {spinmapping}$),
it can been seen that this is an eigenvetor of $\cal H$
with an eigenvalue
\begin{equation}
    \sum_{l=1}^{N_{e}^{v}} \epsilon ^{(v)} (-{p}_{l}) +
     \sum_{m=1}^{N_{e}^{c}} \epsilon ^{(c)} ({q}_{m})
      -\frac{U}{N}\ [\ r^{2}-(N_{{\cal D}_{j}}+1)\ r
        +\frac{N_{e}^{pair}}{2}\ (N_{{\cal D}_{j}}-\frac{N_{e}^{pair}}{2}+1)\ ]
\label {eigenvalue}
\end{equation}
where
$N_{e}^{pair}$ and $r \ (\ 0\le r\le \frac {N_{{\cal D}_{j}}}{2} ,
\ \ r:{\rm integer}\ )$
are defined by
$N_{{\cal D}_{j}}-2S^{z}$ and $\frac{N_{{\cal D}_{j}}}{2}-S\ \ $
respectively.
The total number of the fermions is given by
$N_{e}^{v}+N_{e}^{c}+N_{e}^{pair}$.
Varying the index $j$,
$\cal H$ is diagonalized completely.

Now let us consider the physical properties of the system
in the thermodynamic limit ($N \rightarrow \infty $).
For simplicity,
we consider the half-filled case,
namely, $N_{e}^{v}+N_{e}^{c}+N_{e}^{pair}=N$.
When the interaction is absent,
the system is insulating.

Let us first consider the zero-temperature phase diagrams.
The ground state was obtained by minimizing the energy ($\ref {eigenvalue}$).
The competition
between the kinetic term and the interaction term
gives a rich phase diagram.
We present the phase diagrams of two cases:
the one-dimensional two-band model,
as shown in Fig. 2,
and a system with a constant density of states,
which resembles that of the 2-d  systems ( Fig. 3 ).
We find
three types of different phases
as shown in Figs. 2, 3.
All the phases are separated
by the first-order phase transitions.
The phases are characterized by
$\Delta $
( $=$
the amplitude of the off-diagonal long-range order $\cite {off1}$)
which is defined by
\begin{equation}
\Delta =
\sqrt {\frac{1}{N^{2}}\int dx\int dy \ \langle \psi ^{(c)^{\dagger}}(x)
                                        \psi ^{(v)^{\dagger}}(x)
                                        \psi ^{(v)}(y)
                                        \psi ^{(c)}(y)\rangle }.
\label{odlro}
\end{equation}
Here
$\ \psi ^{(c)}(x)=\frac{1}{\sqrt N}\sum _{k}e^{ikx}c_{k}^{(c)}\ $
and
$\ \psi ^{(v)}(x)=\frac{1}{\sqrt N}\sum _{k}e^{ikx}c_{k}^{(v)}\ $.
The contents of the three phases are as follows;
\begin{flushleft}
$\bf Phase1$:
 $\Delta =0. 5$, which is the upper bound for $\Delta $.
 It is superconducting $\cite {off1}$$\cite {off2}$.

$\bf Phase2$:
 $0<\Delta <0.5$.
 It is also superconducting.

$\bf Phase3$:
 $\Delta =0$.
 The ground state is a band insulator as the non-interacting case.
\end{flushleft}
Note that a sufficiently large attraction is needed
to produce the superconductivity,
which is totally different from the BCS superconductivity.

Now let us discuss the Meissner effect,
namely, estimate the superfluid density $N_{s}$.
$N_{s}$ is defined by ${\frac {mc}{e^{2}}}{\frac {||\ {\bf j}\ ||}
{||{\bf A}||}}$,
where $m$ denotes the effective mass, ${\bf j}$ the current density and
${\bf A}$ the vector potential.
Since we have diagonalized the Hamiltonian explicitly,
it is straightforward to obtain $N_{s}$ by the use of the Kubo formula.
In the insulating phase we obtain $ {N_{s}}/{N}=0$
and there is no Meissner effect.
This is the direct consequence of the effective mass theorem $\cite {mass}$.
In the superconducting phase
we can also obtain ${N_{s}}/{N}=1+O(1/{U})$
in the large $U$ limit, which means the Meissner effect.

Next we consider the thermodynamic properties.
For simplicity,
we consider a system with flat bands
($\epsilon ^{(v)}=-\epsilon $ $\epsilon ^{(c)}=\epsilon $.)
When the two-bands degenerate, namely $\epsilon =0$,
the thermodynamic properties are investigated
by Thouless $\cite {neta3}$.
The grand partition function is
\begin{equation}
{\cal Z}_{\rm ground}
= \sum _{{N_{e}^{v},N_{e}^{c}} \atop {0\le N_{e}^{v}+N_{e}^{c}\le N}}
  \sum _{{r;\ {\rm integar}} \atop {0\le r\le {\frac {N_{{\cal D}_j}}{2}}}}
  \sum_ {N_{e}^{pair}=2r}^{2N_{{\cal D}_j}-2r}
  {\cal C}{\ \ \rm exp}(-\beta {\cal E}),
\label {z}
\end{equation}
where $\ {\cal C}\ $ and $\ {\cal E}\ $ is defined by
$\ {\cal C}=
\frac {N!}{{N_{{\cal D}_{j}}}!\ N_{e}^{v}!\ N_{e}^{c}!\ }
\frac {N_{{\cal D}_{j}}\ !\ (N_{{\cal D}_{j}}-2r+1)}
{r\ !\ (N_{{\cal D}_{j}}-r+1)\ !}\ $
and
$\ {\cal E}=
-\mu (N_{e}^{v}+N_{e}^{c}+N_{e}^{pair})
 -N_{e}^{v}\epsilon+N_{e}^{c}\epsilon
-\frac{U}{N}\{\ r^{2}-(N_{{\cal D}_{j}}+1)\ r
+\frac{N_{e}^{pair}}{2}(N_{{\cal D}_{j}}-\frac{N_{e}^{pair}}{2}+1)\}\ $.
In the thermodynamic limit ($N \rightarrow \infty $)
we use the saddle-point method.
The chemical potential is set $\mu =0$
and the system is half-filling.
A direct calculation leads to analytic forms
of the thermodynamic quantities.
For example, $\Delta (T)$ is given by
\begin{equation}
\Delta ({T})=0.5\ \frac {y^{2}-1}{y^{2}+ay+1},
\label {delta}
\end{equation}
where
$y$ is the largest root of
$\ \log x=\frac{1}{2}\ UT^{-1}(x-1)(x+1)(x^2+ax+1)^{-1}$
and
$\ a=e^{{\epsilon }/{T}}+e^{-{\epsilon }/{T}}$.
As shown in Fig. 4,
the second-order phase transition occurs at a finite temperature.
The critical temperature $T_{c}$
is proportional to $U$ when $U \gg \epsilon $.
The entropy $S(T)$ per unit cell is given by
\begin{eqnarray}
S({T})/N=\log (y^2+ay+1)
-\epsilon \ T^{-1}
\frac {(e^{\epsilon /T}-
e^{-\epsilon /T})y}{y^2+ay+1}
\nonumber
\\
-0.5\ U\ T^{-1} \frac {y(2y+a)(y^2-1)}{(y^2+ay+1)^2}.
\label {entropy}
\end{eqnarray}
The heat capacity ($=T(\partial S/\partial T)_{V}$)
per unit cell is shown in Fig. 4.
In the superconducting phase
it behaves as $A\exp (-\frac {B}{T})$ at a sufficiently low temperature,
where $A$ is a constant and
$B=\frac {U}{2}-2\epsilon$ is the excitation gap.
In the high-temperature phase
it is a decreasing function of $T$,
since the band widths are finite.

We find that
$\frac {\Delta (T=0)}{T_{c}}$ and $\frac {\Delta C}{C_{n}}$
are not universal in contrast to the BCS-type superconductivity,
where $\Delta C$ is the jump of the heat capacity at $T=T_{c}$
and $C_{n}$ is the heat capacity at $T=T_{c}+0$.
A more detailed study
of the thermodynamic properties
will be presented elsewhere.

The half-filled case considered here seems to be most prospective
to be realized.
The crucial point is the origin of the attractive interaction.
One of the possible candidates is the exciton mechanism
proposed in Ref.$\cite {KT2}$ and $\cite {Little}$.
There the attraction is envisaged
as arising from a polarizable medium
sandwiched between the two chains,
where the ``effective'' interaction between electrons
in different chains becomes statically attractive.
This is because electrons share positive charge
induced in the medium.
They have confirmed
that there are cases in which this attractive interaction
is stronger than the direct Coulomb repulsion
between electrons in different chains.
( In Ref.$\cite {AK}$, another example of attraction was proposed
in the two-band repulsive Hubbard model.
The electrons in one band experience attractive interaction
mediated by an accompanying Mott-insulator band ).
Then, if we consider the filling other than half-filling,
the exciton-electron interaction
which leads to the attraction is reduced considerably by screening.
Thus the half-filling case is best for our purpose.
Without the screening, a strong attraction is rather easily achieved
$\cite {attraction}$.

In summary, a recent proposal
by Kohmoto and Takada of the new pairing state
between a conduction electron and a valence electron
was investigated through a family of exactly solvable models.
We obtained all the eigenvalues and the eigenvectors explicitly.
The zero-temperature phase diagrams were obtained.
The superconducting instability without a Fermi surface
which was proposed by Kohmoto and Takada were confirmed.
It was also proved that a sufficiently large attraction
between states in the two bands
is needed to produce supercondutivity.
The thermodynamic properties were also dicussed.
The properties are quite different
from those of the BCS-type superconductor.
The models we consider may be realized
in specially synthesized double-chain organic materials.
Although we have presented the results for the cases
where fully analytical treatments are possible,
the results for the more general cases
are not different from the present cases in essential ways.
They will be presented elsewhere.

\begin{figure}

Fig. 1.

The classification of the pairs and
mapping to a quantum spin.

Fig. 2.

The one-dimensional two-band model,
where
$\epsilon ^{(c)}(k)=-2t\cos k+2t+G/2$
and
$\epsilon ^{(v)}(k)=2t\cos k-2t-G/2$.
The zero-temperature phase diagram.

Fig. 3.

The model which has a constant density of states.
The density of states  and the zero-temperature phase diagram.

Fig. 4.

The temperature dependence
of the order parameter and the heat capacity
when $\epsilon =0.3$ and $U=2$.

\end{figure}

\end{document}